# Real-time simulation of large-scale HTS systems: multi-scale and homogeneous models using *T-A* formulation

Edgar Berrospe-Juarez[1], Víctor M R Zermeño[2], Frederic Trillaud[3] and Francesco Grilli[4]

[1]Postgraduate School of Engineering, National Autonomous University of Mexico, Mexico

[2]NKT

[3]Institute of Engineering, National Autonomous University of Mexico, Mexico

[2]Karlsruhe Institute of Technology, Germany

email: eberrospej@iingen.unam.mx

This work was supported in part by the Programa de Maestría y Doctorado en Ingeniería of the Universidad National Autónoma de México (UNAM) and the Consejo Nacional de Ciencia y Tecnología (CONACYT) under CVU: 490544, and by DGAPA-UNAM grant, PAPIIT-2017 #TA100617.

**Abstract:** The emergence of second-generation high temperature superconducting tapes has favored the development of large-scale superconductor systems. The mathematical models capable of estimating electromagnetic quantities in superconductors have evolved from simple analytical models to complex numerical models. The available analytical models are limited to the analysis of single wires or infinite arrays that, in general, do not represent actual devices in real applications. The numerical models based on finite element method using the *H* formulation of the Maxwell's equations are useful for the analysis of medium-size systems, but their application in large-scale systems is problematic due to the excessive computational cost in terms of memory and computation time. Then it is necessary to devise new strategies to make the computation more efficient. The homogenization and the multi-scale methods have successfully simplified the description of the systems allowing the study of large-scale systems. Also, efficient calculations have been recently achieved using the *T-A* formulation. In the present work, we propose a series of adaptations to the multi-scale and homogenization methods so that they can be efficiently used in conjunction with the *T-A* formulation to compute the distribution of current density and hysteresis losses in the superconducting layer of superconducting tapes. The computation time and the amount of memory are substantially reduced up to a point that it is possible to achieve real-time simulations of HTS large-scale systems under slow ramping cycles of practical importance on personal computers.

# 1. Introduction

In recent years, important progress has been achieved in the manufacturing of high current, high temperature superconductors (HTS). Nowadays, the high current capacity of second-generation (2G) HTS tapes has made possible the production and the commercialization of superconducting fault current limiters and cables. The interest in the technology has spawned over to other large-scale superconducting systems such as generators, motors [1], and high field scientific magnets [2-4]. The design and operation of those systems are challenged by the difficulty of extracting the heat produced by the hysteresis losses. In this manuscript, we are using the term hysteresis losses instead of AC losses, because the AC losses may include other contributions like the losses in the copper and other normal conductors [5]. However, in the present work, we are only focusing our attention to the losses in the superconductor material. Then, the methods proposed here allow the accurate estimation of the current density in the superconductor material and the magnetic field over the entire system. This degree of accuracy enables not only to achieve good results on the losses but also to address the study of systems where high precision in the magnetic field is required, like MRI and NMR systems.

The first available analytical models were limited to the analysis of one single conductor [6-8]. The disagreement between the analytical calculations and the experimental measurements as well as the importance of the electromagnetic interaction between conductors were first discussed in [9]. More complex models considering conductor stacks under particular conditions are presented in [10-13]. The reader can refer to [14] for a thorough review of existing analytical models. On the other hand, numerical models opened up the possibility of considering systems with larger number of tapes. Methods like the Finite Element Method (FEM) are well documented in the literature [15-17] and have been extensively applied to superconductor systems. Nevertheless, they have specific characteristics that should be considered in order to successfully address the modelling task [18-19]. For a deeper review of numerical models, the reader is referred to [5, 20]. Among the available methodologies suitable for the analysis of large-scale HTS systems, one should also mention the Minimum Magnetic Energy Variation (MMEV) method introduced in [21-22]. The MMEV method has then evolved into the Minimum Electro-Magnetic Entropy Production (MEMEP) method [23-24].

The FEM models of HTS tapes use different formulations of Maxwell's equations. The differences between the formulations come from the variety in the choice of the state variables. The first proposed FEM model of an HTS tape used the $T$-$\phi$ formulation [25], a later model applied the $A$-$V$ formulation [26]. Other formulations are listed in [19] and [27]. During the last years the $H$ formulation [26] has been widely used in the community. In two recent publications [28-29], the $T$-$A$ formulation has been proposed as an efficient approach to simulate thin layers of superconductor. This formulation couples the $T$ and $A$ formulations. Thus far, it appears as one of the most efficient approaches to address the analysis of large-scale HTS systems.

The main concern when dealing with large-scale HTS systems is the large number of turns. The homogeneous and the multi-scale approaches can considerably reduce the computational cost of the models without compromising accuracy. The homogenization process assumes that a stack of tapes can be represented by a single homogeneous anisotropic bulk. This approach was first proposed in [30] and later improved in [31-33]. The multi-scale approach is based on the idea of simulating a subset of tapes, referred to as analyzed tapes. The interaction between a single tape and the full model made of all the tapes is achieved by computing the magnetic field generated by all the tapes. The resulting magnetic field is subsequently used as boundary condition to solve the current density in the analyzed tapes. This method was proposed in [34], with later refinements added in [20]. The last version of the method incorporates an iterative process to improve the accuracy of the current density in the analyzed tapes and, consequently, of the losses [35].

In the present work, two strategies, multi-scale and homogeneous, used together with the *T-A* formulation are presented to address the numerical modelling of large-scale HTS systems. To couple those strategies with the proposed *T-A* formulation, some specific considerations are required. Thus, the proposed strategies apply only to conductors that can be approximated by a line (1D elements) such as the 2G HTS tapes, for instance. They assume that the current density distribution of the full the system can be estimated by either considering a reduced number of analyzed tapes, in the case of the multi-scale strategy, or a bulk instead of a stack, in the case of the homogenization. It is demonstrated that the proposed strategies can achieve real-time simulations of real charge/discharge cycles of HTS magnets on a personal computer.

This paper is organized as follows: Section 2 makes a brief presentation of the *T-A* formulation [28-29]. Section 3 contains the description of the case study, a racetrack coil made of 2,000 turns. This case study and its *H* formulation model are used to validate the *T-A* formulation model as well as the choice of the order of the elements. The description of new proposed *T-A* multi-scale and homogeneous models is presented in section 4. Section 5 contains the validation of the multi-scale and homogenous models by comparing them with the previously validated *T-A* full model, the results demonstrate the real-time capabilities of the proposed models. The conclusions are exposed in section 6. In the appendices, we present different side studies that arose during the implementation and the use of the *T-A* formulation and its comparison with the well-known *H* formulation. These studies are not the main focus of the work but they are certainly of interest to the applied superconductivity community. In the appendix A, we present the analysis of the order of the elements used in the *T-A* formulation to find a stable pair of finite element spaces. The analysis is conducted by means of performing different simulations using different order of elements to approximate **T** and **A**. Appendix B is also included to show the computation time limitations of the *H* formulation models.

## 2. *T-A* formulation

Hereinafter, we briefly recall some of the salient information of the *T-A* formulation. For further information, the reader is referred to [28, 29]. The new proposed multi-scale and homogenization strategies allow simulating real-time slow charge and discharge of large-scale HTS systems while keeping a high accuracy in the current distribution in each of its individual tape. These new strategies are going to be described in detail in section 4. It is recalled that the *T-A* formulation relies on the primary assumption that thin superconductor layers of the HTS tapes can be modelled as one dimensional (1D) elements. Therefore, it is assumed that the bounded universe is made of 1D superconducting layers surrounded by a homogeneous, isotropic medium (metallic layers, insulation and the cryogenic liquid) referred to as surrounding medium. The current vector potential **T** is exclusively defined over the superconducting layers, and the magnetic vector potential **A** is defined over the entire bounded universe. The corresponding boundary conditions are applied to the vector **A** at the boundary of the universe. The current is assumed to circulate only in the superconducting layers and the rest of the universe is considered non-conductive.

The governing equation of the *A* formulation is

$$\nabla^2 \mathbf{A} = -\mu \mathbf{J}, \tag{1}$$

where $\mu$ is the magnetic permeability, $\mathbf{J}$ the current density, and **A** is defined as $\mathbf{B} = \nabla \times \mathbf{A}$, where **B** is the magnetic flux density [27].

The governing equation of the *T* formulation is

$$\nabla \times \rho \nabla \times \mathbf{T} = -\frac{\partial \mathbf{B}}{\partial t}, \tag{2}$$

where $\rho$ is the resistivity and **T** is given by $\mathbf{J} = \nabla \times \mathbf{T}$ [27].

In the 2D case depicted in figure 1, the thickness of the superconducting layers is neglected, and **T** has a single scalar component $T_y$, that leads to simplify equation (2) to

$$\frac{\partial}{\partial x}\left(\rho_{HTS} \frac{\partial T_y}{\partial x}\right) = \frac{\partial B_y}{\partial t}, \tag{3}$$

where $\rho_{HTS}$ is the resistivity of the HTS material. The magnetic field component $B_y$ is obtained by calculating $A_z$. The current density has only one component defined by $J_z = \partial T_y/\partial x$. The necessary boundary conditions at the edges of the 1D superconducting layer for **T** can be obtained by integrating the current density **J** over the cross-section of the layer which is equal to the transport current in the tape, as follows,

$$I = \iint_S \mathbf{J}\, dS = \iint_S \nabla \times \mathbf{T}\, dS = \oint_{\partial S} \mathbf{T}\, dr, \tag{4}$$

where $S$ is the cross-section of the superconductor, and $\partial S$ is its boundary. As shown in figure 1, the component of **T** parallel to the layer is zero, then (4) takes the following expression,

$$I = (T_1 - T_2)\delta \tag{5}$$

where $\delta$ is the real thickness of the HTS layer. $T_1$ and $T_2$ are the potentials at the extremities of the 1D layer as shown in figure 1. Thus, a different transport current can be impressed by modifying $T_1$ and $T_2$.

The current density is multiplied by $\delta$, to obtain a surface current density $K = J_z \cdot \delta$. The surface current density $K$ is impressed into the *A* formulation as an external surface current density by means of a boundary condition of the form

$$\mathbf{n} \times (\mathbf{H}_1 - \mathbf{H}_2) = \mathbf{K} \tag{6}$$

where **n** is the unit vector normal to the tape, while $\mathbf{H}_1$ and $\mathbf{H}_2$ are the magnetic field strength vectors above and below the layer, respectively.

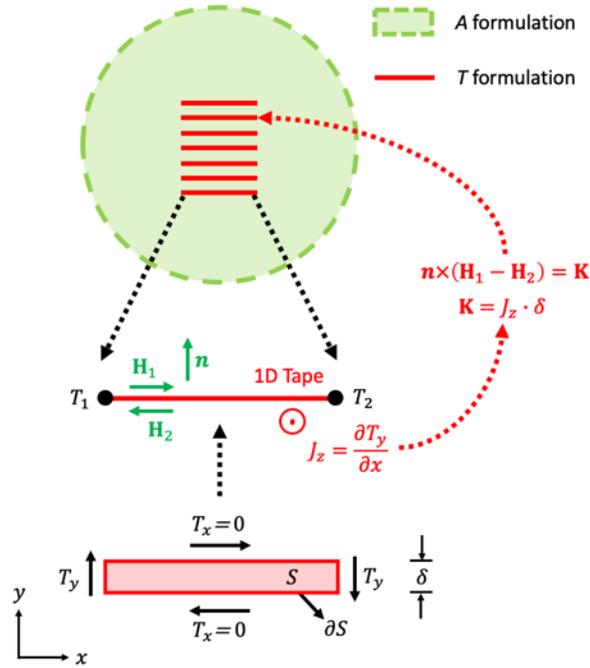

**Figure 1.** Bounded universe with superconductive layers and surrounding medium. The current vector potential **T** is computed over the HTS layers and the magnetic vector potential **A** is computed over the entire bounded universe. The surface current **K** is impressed in by means of a boundary condition. The rectangle of surface $S$ is used in this figure to show how the boundary conditions are deduced.

As mentioned previously, the 1D assumption restricts the application of the *T-A* formulation to systems where the superconducting layers have a large aspect ratio, as in the 2G HTS tapes. The approximated current density can only vary along the 1D line, then it is not possible to consider the penetration of the parallel magnetic field component into the tape. However, the influence of both the parallel and perpendicular components of the magnetic field are taken into account for the purpose of computing the critical current density ($J_c$). As a consequence, the *T-A* formulation is only suitable for cases where the influence of the parallel component of the field is negligible, examples of these kind of cases are the coils analyzed in [20, 28, 29]. The cases where the influence of the parallel component cannot be neglected, like the long solenoid presented in [36], cannot be addressed with this formulation.

## 3. Case study and full models

### *H* full model

The case study used in this manuscript is the same racetrack coil used in [20]. The coil has 10 pancakes (stacks), each composed of 200 turns. The symmetry allows to model just one quarter of the system, this means it is possible to consider 5 pancakes each one with 100 tapes (500 tapes in total).

As mentioned in section 1, there exist different approaches to model HTS systems, some of them are analytical, some others are numerical; each approach has its own limitations and area of application. The comparison of the new proposed strategies against all the suitable approaches is beyond the scope of this work. The *H* formulation models are one of the most widespread approaches and have been validated against experimental data [37]. Hence, in this section, we consider the reference model used in [20] as the best possible model, and we use it to validate the new proposed strategies. This model uses the *H* formulation considering each tape, and is called *H* full model.

We refer to the regions containing just one HTS layer and its surrounding medium as unit cells. The mesh of the unit cells is structured, and considers 1 element along the tape's thickness and 100 elements along its width. The 100 elements are distributed symmetrically with respect to the center of the tape with an increasing number of elements at the extremities of the tape, see figure 2. Since the resistivity of the surrounding medium is several orders of magnitude larger than the resistivity of the superconducting layer, the losses in the normal conductors forming the HTS tapes are negligible compared to the losses in the superconductor. Thus, the presence of those materials is neglected and the HTS layers are considered to be surrounded with a medium having a resistivity value $\rho_a = 1\ \Omega\text{m}$, as proposed in [33].

The electrical resistivity of the HTS material is modeled by the so-called *E-J* power-law [38],

$$\rho_{HTS} = \frac{E_c}{J_c(\mathbf{B})} \left| \frac{\mathbf{J}}{J_c(\mathbf{B})} \right|^{N-1}. \tag{7}$$

The critical current density $J_c$ is defined by a modified Kim's relation [39], this relation describes the anisotropic dependence of the critical current density on the magnetic flux density as follows

$$J_C(\mathbf{B}) = \frac{J_{c0}}{\left(1 + \frac{\sqrt{k^2 B_\parallel^2 + B_\perp^2}}{B_0}\right)^\alpha}, \tag{8}$$

where $B_\perp$ and $B_\parallel$ are the magnetic field components perpendicular and parallel to the wide surface of the tape. The magnetic field considered in (7) and (8) includes both background field and self-field. Since there are no magnetic materials, the permeability of the surrounding medium and the HTS material is chosen to be the permeability of the vacuum $\mu_0$. The parameters of racetrack coil are summarized in table 1.

**Table 1.** Case study coil parameters.

| Parameter | Value |
|---|---|
| Pancakes | 10 |
| Turns per pancake | 200 |
| Unit cell width | 4.45 mm |
| Unit cell thickness | 293 μm |
| HTS layer width | 4 mm |
| HTS layer thickness | 1 μm |
| $E_c$ | 1e-4 Vm$^{-1}$ |
| $N$ | 38 |
| $J_{c0}$ | 2.8e10 Am$^{-2}$ |
| $B_0$ | 0.04265 T |
| $k$ | 0.29515 |
| $\alpha$ | 0.7 |
| Surrounding medium $\rho_a$ | 1 Ωm |

## *T-A* full model

The *T-A full model* is a model that considers all the tapes, this means that **T** is computed in every tape. The *T-A* full model uses first order elements for **T** and second order elements for **A**, this election is justified in the appendix A. The mesh used in the unit cells is a structured mesh similar to the *H* full model. The geometry and the mesh of the model is shown in figure 2. This figure also shows the numbering of the tapes and pancakes.

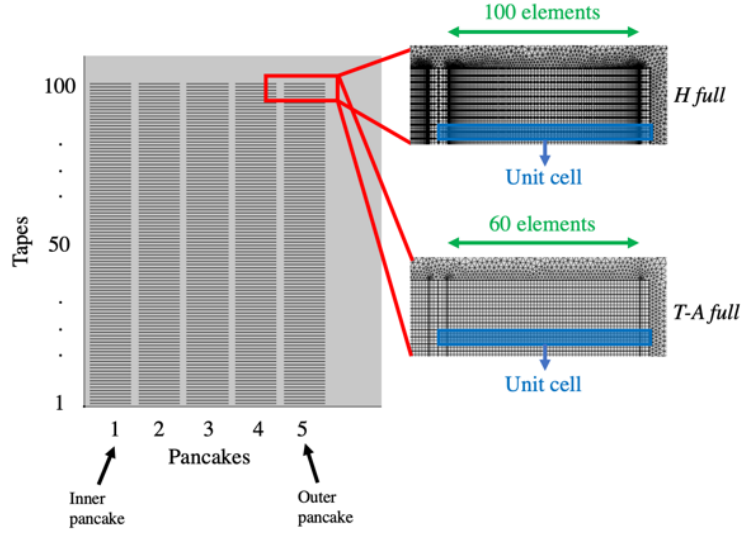

**Figure 2.** Geometry and mesh of the *H* full model and the *T-A* full model. The analyzed section of the case study has 5 pancakes with 100 tapes per pancake. The mesh in the unit cells is structured elements with 100 elements along the tape's width for the *H* full model and 60 elements for the *T-A* full model.

**Numbers of elements along the tape's width**

To validate the *T-A* full model, the *H* full model and the *T-A* full model were simulated for one cycle of a 11 A, 50 Hz transport current. The value 11 A was chosen because at this current value the tape 1 in pancake 5 is completely penetrated by transport and magnetization currents. To assess the impact of the numbers of elements along the tape's width, the *T-A* full model was modified to consider different numbers of elements ranging from 25 to 150. The distribution of the elements along the tape's width is uniform. It should be recalled that in the *H* full reference model an increasing number of elements is considered at the extremities of the tapes. That is why this assessment goes beyond the 100 elements considered in the *H* full model. All the models discussed in this manuscript were implemented in COMSOL Multiphysics 5.3, and the personal computer used to perform the simulations was a MacBookPro (3 GHz Intel Core i7-4578U, 4 cores, 16 GB of RAM).

The quantitative comparison of the *H* and *T-A* full models is carried out by calculating the relative error of the average hysteresis losses and the coefficient of determination ($R^2$) of the *J* distributions. The average hysteresis losses were obtained considering data of the second half of the cycle, as follows,

$$Q_{av} = \frac{2}{P}\int_{P/2}^{P}\int_{\Omega} E \cdot J \, d\Omega \, dt, \qquad (9)$$

where *P* is the period of the sinusoidal cycle, and $\Omega$ is the superconductor domain.

The relative error on the losses is defined as $er_q = |(Q_{H\_av} - Q_{TA\_av})/Q_{H\_av}| \times 100\%$, where $Q_{H\_av}$ and $Q_{TA\_av}$ are the losses computed with the $H$ and $T$-$A$ full models, respectively. The coefficient of determination is a widely used metric to evaluate the goodness of a fit [40], here it is used to compare the shape and evolution of the $J$ distributions across the tape width over space and time. Indeed, they are multivariable functions and not just mere scalars. The coefficient of determination is defined as,

$$R^2 = 1 - \frac{\sum_{i=1}^{m}(J_H - J_{TA})^2}{\sum_{i=1}^{m}(J_H - \bar{J}_H)^2} \ , \tag{10}$$

where $J_H$ and $J_{TA}$ are vectors containing the uniformly sampled $J$ distributions over the tape width of all the tapes, for all the time steps, computed with the $H$ and $T$-$A$ models, respectively. $\bar{J}_H$ is defined as the average of $J_H$. It must be remembered that $R^2 = 1$, means a perfect matching between $J_H$ and $J_{TA}$.

The results of the simulations using different number of elements are resumed in figure 3. The losses relative error and $R^2$ are plotted as a function of the number of elements along the tape's width. The lesser elements in the tapes, the larger the losses error, and the lower the $R^2$. As the number of elements is incremented the relative error $er_q$ and $R^2$ present an asymptotic behavior. An extra axis is included in the upper part of the plot to report the computation time, it can be observed that the increment in the number of elements produces an increment in the computation time, and when the number of elements is 150 the computation time is more than the time required by the $H$ full model. We consider that the compromise between accuracy and computation time is fulfilled with 60 elements along the tape's width, because the error of the corresponding losses is less than 1 % and the $R^2$ is more than 0.99. Throughout the rest of this manuscript it is assumed that the $T$-$A$ model has 60 elements along the tape's width.

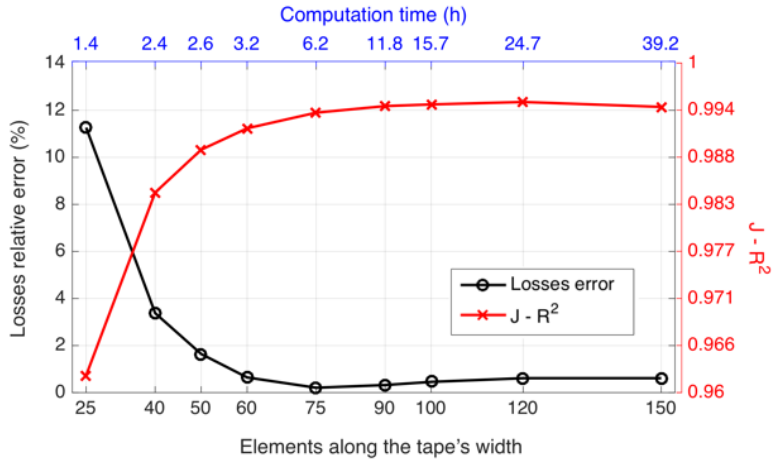

**Figure 3.** Losses relative error ($er_q$) and coefficient of determination ($R^2$) of the $J$ distribution as a function of the number of elements along the tape's width. The accuracy improves as the number of elements increases, but both parameters show an asymptotic behavior. The upper axis shows the computation time.

The $|\mathbf{B}|$ at the peak of transport current ($t = 15$ ms) computed with both $H$ and $T$-$A$ full models are shown in the first row of figure 4. The normalized current density ($J_n = J/J_c$), also at peak current, is presented in the second row. The third row in figure 4 shows the average hysteresis losses in the coil, following the numbering presented in figure 2. The x-axis in the plots represents the tape's number. There are five lines, each one representing a different pancake. The losses in pancake 5 are almost three orders of magnitude larger than the losses in pancake 1. This difference is linked to the increase in the current penetration in the tapes moving from pancake 1 to pancake 5. Also, the higher losses in pancake 5 are related to the difference in the field direction, having a higher perpendicular component. The plots in figure 5 demonstrate that the $T$-$A$ full model can achieve accurate estimations of the electromagnetic quantities at a local level, i.e. along each individual tape. As a result, the losses are nearly identical across the coil.

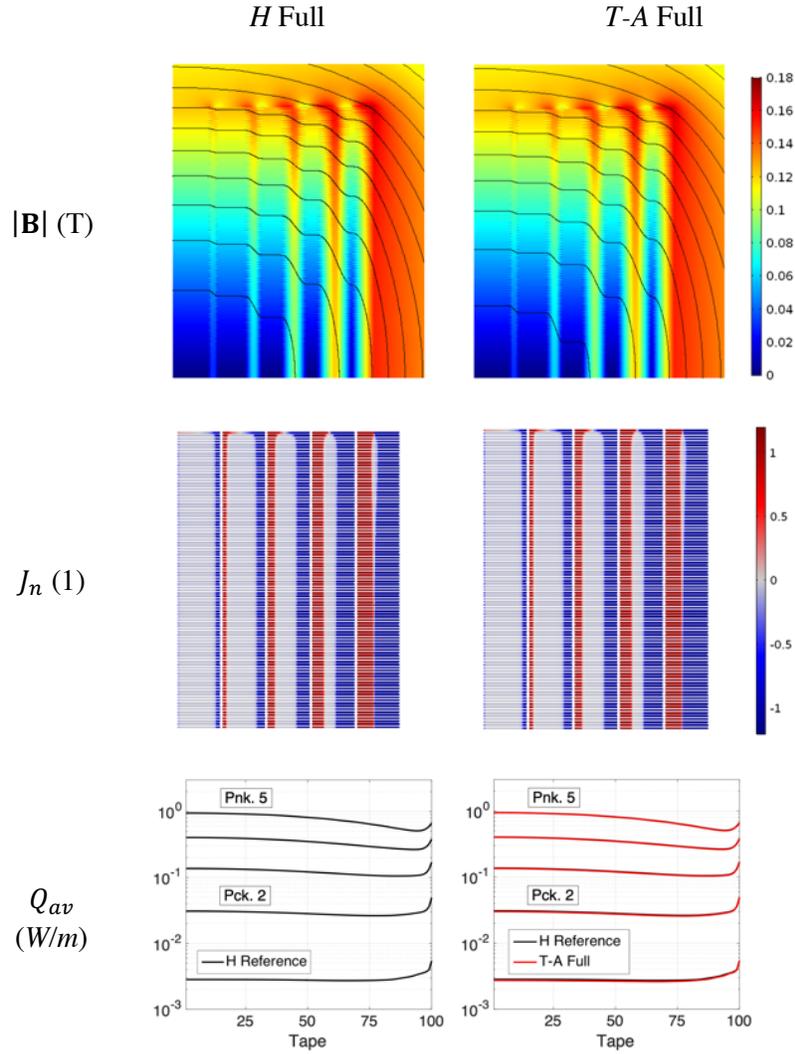

**Figure 4.** Results of *H* full and *T-A* full models. The first and second rows show |**B**| and $J_n$ at peak transport current ($t$ = 15 ms). The third row shows the average losses as a function of the tape's number inside each pancake. The *T-A* model uses first order elements for **T** and second order elements for **A**, as well as 60 elements along the tape's width.

## 4. *T-A* multi-scale and homogeneous approaches

This section presents two refined strategies, based on the *T-A* formulation. These strategies make use of less resources, compared with the *T-A* full models, while obtaining similar accuracies for the hysteresis losses as well as current density and magnetic field distributions.

## *T-A* multi-scale approach

The multi-scale method, as presented in [20, 35], uses two different submodels, single tape and coil. The single tape submodel uses the *H* formulation and computes the *J* distribution in the analyzed tapes. The coil submodel uses the *A* formulation and computes the magnetic field throughout the whole system. The multi-scale technique developed here lumps the two submodels into a single model using the *T-A* formulation. Hence, this new multi-scale model is able to compute simultaneously the *J* distribution in the analyzed tapes and the background field produced by all the tapes.

In the multi-scale *T-A* model, **T** has to be defined for each individual analyzed tape. The *J* distribution along each analyzed tape is obtained by calculating **T**, and the *J* distributions in the non-analyzed tapes are approximated by linear interpolation from neighboring analyzed tapes. The potential **A** is defined over the entire bounded universe and its computation in turn allows the computation of **B**. The *J* in the analyzed and non-analyzed tapes is multiplied by the thickness of the superconducting layer ($\delta$), to obtain a surface current density (**K**) which is impressed into the *A* formulation by means of a boundary condition of the form (6). Figure 5 shows the multi-scale approach applied to a small stack.

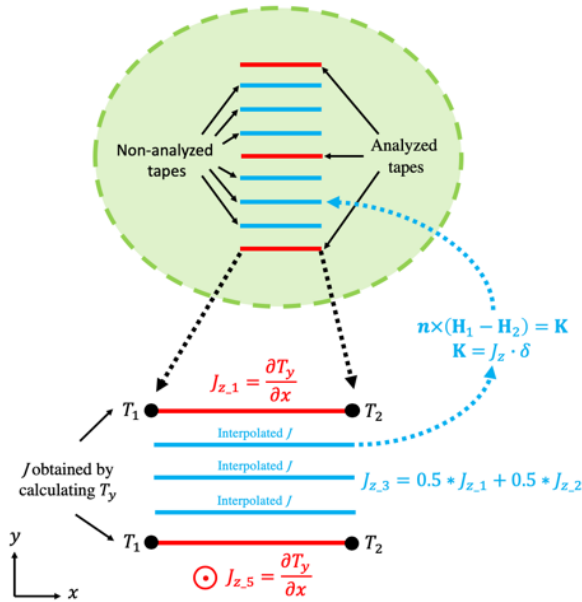

**Figure 5.** *T-A* multi-scale approach. In this example, there are 3 analyzed tapes whereby *J* is obtained by computing **T**. The *J* in the remaining non-analyzed tapes is approximated by linear interpolation. The surface current **K** is impressed by means of a boundary condition.

## *T-A* homogenous approach

The homogenization technique assumes that a stack of HTS tapes can be modeled as a homogeneous anisotropic bulk [33]. This simplification in the geometrical description should not compromise the electromagnetic behavior. The transformation of a small stack into a bulk that brings together all the unit cells containing the HTS tapes is depicted in figure 6.

Once again, **A** is defined all over the entire bounded universe, while **T** is now only defined inside the bulk. Inside this bulk, for the purpose of computing **T**, the influence of the component of **B** parallel to the surface of the tapes is not considered. Therefore, **T** is forced to have only one component. In the example showed in figure 6, $T_y$ is the only component of **T** and is defined by means of equation (3).

The bulk can be understood as the limiting case of a densely packed stack made up of 1D superconducting layers. Each of them should transport the same current of its original counterpart, which in turn requires the boundary conditions $T_1$ and $T_2$, defined in equation (5), to be applied to each 1D layer. These boundary conditions should be applied to the corresponding edges of the bulk, as depicted in figure 6. Additionally, the boundary conditions in the upper and lower edges of the bulk in figure 6 are Neumann boundary conditions of the form

$$\frac{\partial T_y}{\partial \boldsymbol{n}} = 0 \tag{11}$$

where $\boldsymbol{n}$ represents the unit vector normal to the edge.

The resistivity inside the bulk, for the purpose of computing $T_y$, is considered to be the resistivity of the superconductor material. Thus, the copper and other normal conductors that constitute the HTS tapes are not considered.

The current density inside the bulk has one component defined by $J_z = \partial T_y / \partial x$. A scaled current density is defined as

$$J_s = \frac{\delta}{\Lambda} J_z \tag{12}$$

where $\delta$ is the thickness of a tape and $\Lambda$ is the thickness of the unit cell. The $J_s$, for the purpose of computing $A_z$, is impressed in the bulk domain as a source term as follows

$$\nabla^2 A_z = -\mu_0 (\sigma_0 E_z + J_s) \tag{13}$$

where $\sigma_0 = 0$ is the conductivity of the surrounding medium, which for the purpose of computing $A_z$, is considered as the conductivity of the entire bounded universe.

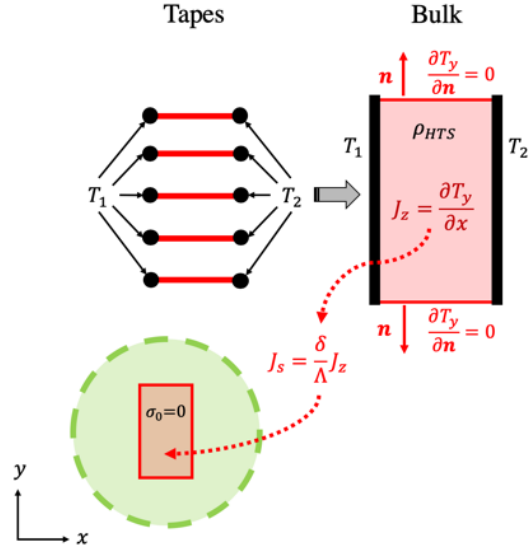

**Figure 6.** *T-A* homogeneous approach. The stack is transformed in a bulk. The influence of the component of $B_x$ is neglected, the $T_y$ is the only component of **T**. The boundary conditions $T_1$ and $T_2$ should be applied to the edges corresponding to the extremities of the tapes. The scaled current density $J_s$ is impressed as a source term.

**Case study *T-A* multi-scale model**

The *T-A* multi-scale model considers 30 analyzed tapes. The election of the analyzed tapes follows the same directives proposed in [35], more analyzed tapes are used in the upper part of the pancakes where the results of the full models present higher variations in the losses. The set of analyzed tapes in each pancake is {25, 66, 88, 96, 99, 100}. This distribution allows reproducing the expected variations in the losses at the upper part of the pancakes.

To avoid oscillations in the *J* distribution, as discussed in the appendix A, the unit cells of the analyzed tapes and their closest non-analyzed tapes should use second order elements to approximate **A**, while first order elements are used to approximate **A** throughout the rest of the system. The domains using first and second order elements for **A** are connected with each other by means of a Dirichlet boundary condition. The analyzed tapes and the domains using second order elements for **A** are shown in figure 7. It is important to mention that first order elements are used for **T**. The mesh of the unit cells is structured, the mesh of the analyzed tapes and their closest non-analyzed tapes consider 60 elements along the tape's width, while 30 elements are considered along the rest of the tapes. The distribution of the elements along the tape's width is uniform. A part of the mesh presenting these features is also shown in figure 7. The use of first order elements for **A** in the non-analyzed tapes as well as less elements along these tapes allows to build a model with less degree of freedom (DOF) without compromising the accuracy of the results.

Once the hysteresis losses are computed along the analyzed tapes, the Piecewise Cubic Hermite Interpolating Polynomial (PCHIP) method [41] is used to approximate the losses in the non-analyzed tapes.

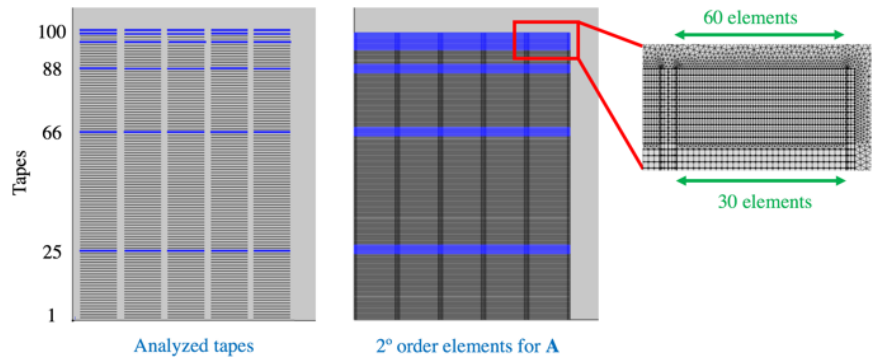

**Figure 7**. Analyzed tapes, regions with second order elements (blue unit cells) and the mesh of the *T-A* multi-scale model. An increased number of analyzed tapes is defined in the upper part of the pancakes. The analyzed tapes and their closet non-analyzed ones use second order elements for **A**. The mesh considers 60 and 30 elements along the tapes.

**Case study *T-A* homogenous model**

The model considers 5 bulks, one for each pancake. The mesh in the bulk's domains is structured considering 6 unequal elements along the bulk's width and 60 elements along the tape's width as shown in figure 8. The distribution of the elements along the bulk's width is similar to the distribution of the analyzed tapes in the multi-scale model; more elements are used in the upper part of the pancakes where the results of the full models present higher variations in the losses. Again, first order elements are used to approximate **T** and second order elements for **A**.

The hysteresis losses are computed along a line at the center of each of the 6 elements making the bulks. These losses, corresponding to the tapes located at the center of the 6 elements, are used to approximate the losses along the rest of the tapes of the pancakes by means of the PCHIP interpolation method.

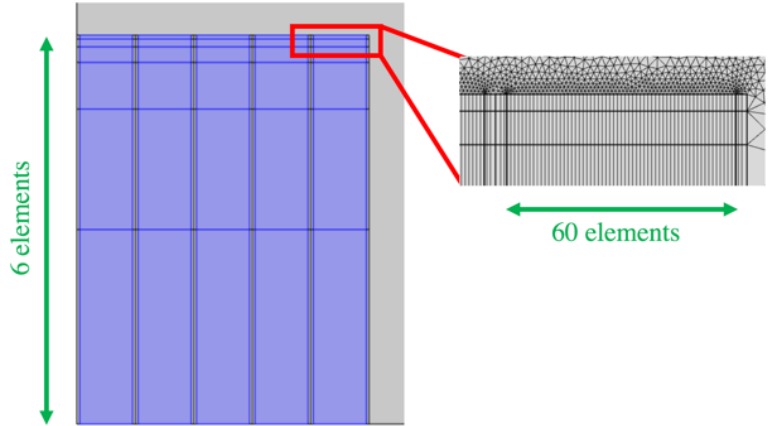

**Figure 8.** Geometry and mesh of the *T-A* homogeneous model. The bulk's mesh considers 6 elements along the bulks, with a denser distribution of elements in the upper part of the pancakes.

## 5. Real-time simulations

The real operation of large-scale HTS systems like high field magnets involves slow charge/discharge cycles, with periods in the order of hours [4]. Considering such slow operations, it is shown that real-time simulations can be achieved using the *T-A* multi-scale and *T-A* homogenous models. Real-time simulation refers to a simulation in which the time required to complete the computation of the results is shorter than the time for the physical phenomenon to occur [42].

*T-A* full model has been validated against the *H* full model in section 3. From the analysis presented in the appendix B, assuming typical solver settings for available solvers in COMSOL, the computation time required to simulate a slow charge/discharge cycle using the *H* full model becomes prohibitive compared to the *T-A* full model. Therefore, the *T-A* full model is now used as reference model for the rest of the study. The *T-A* multi-scale and homogeneous models were simulated for the 11 A charge/discharge cycle showed in blue line in figure 9. The charge/discharge cycle is 1 h ramping up, plus 30 min plateau at 11 A, plus 1 h ramping down, plus 30 min plateau. The losses, computed with the three *T-A*, as a function of time are also shown in figure 9.

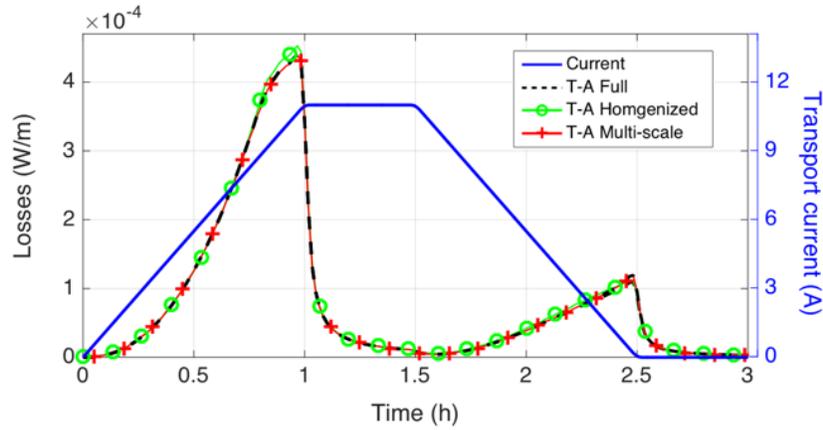

**Figure 19.** Transport current during the slow charge/discharge cycle. Losses as a function of time estimated by the three *T-A* full, multi-scale and homogenous models.

Figure 10 presents the results on the magnetic flux density $|\mathbf{B}|$, current density $J_n$ and the losses $Q$. It allows making a qualitative comparison between simulation results. The magnetic flux density and the current density are computed at $t = 1.5$ h. It can be seen that the *T-A* multi-scale and homogenous models can successfully reproduce the $J$ distribution produced by the *T-A* full model. The $J_n$ plot of the *T-A* homogeneous model shows the $J_n$ all across the bulks. As a direct consequence of the accurate estimation of the $J$ distributions accomplished by the multi-scale and the homogenous models, the magnetic field plots are visually identical, which constitutes a first validation of both approaches. The third row in figure 10 shows the hysteresis losses integrated over the whole charge/discharge cycle.

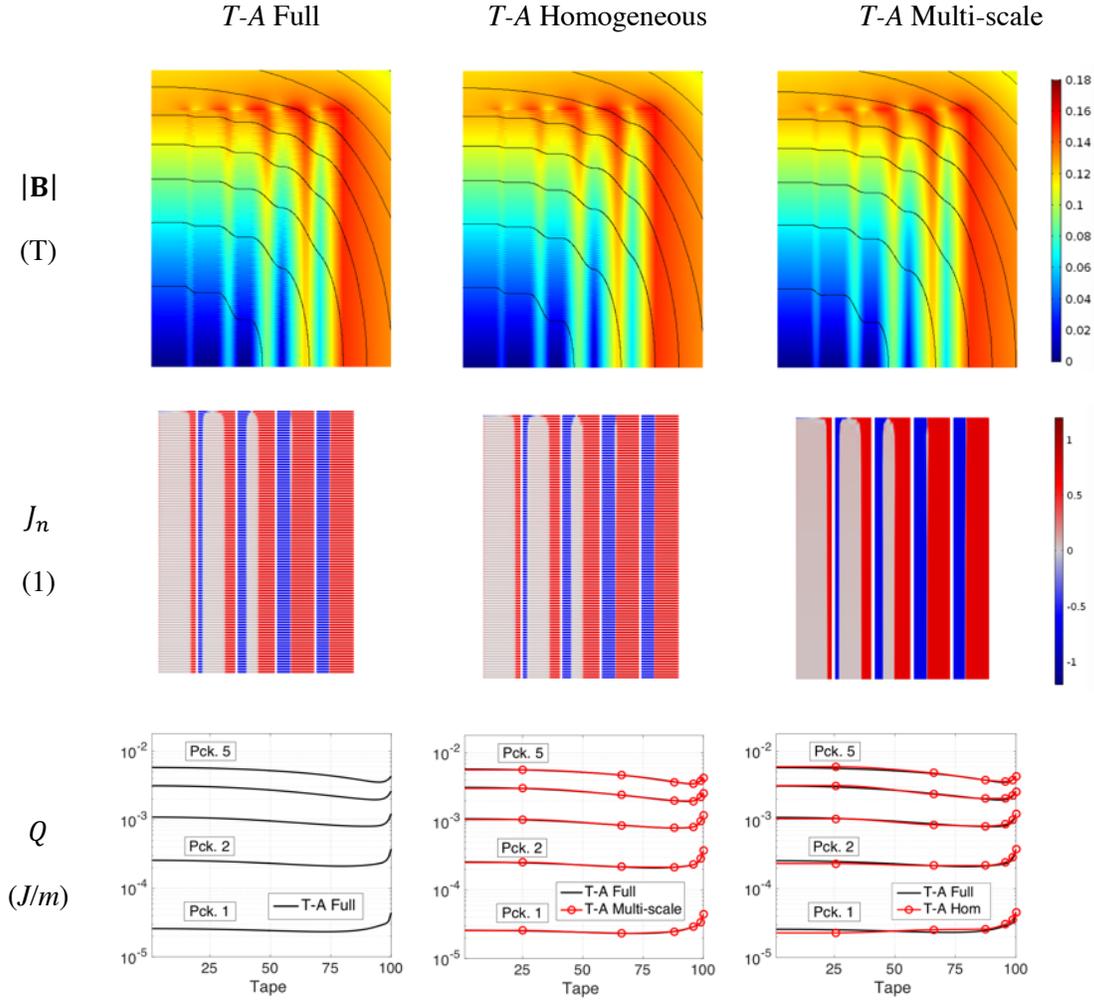

**Figure 10.** Results of the *T-A* full, multi-scale and homogenous models with a charge/discharge transport current. The first column shows the results of the *T-A* full model. The second and third column show the results of *T-A* multi-scale and homogeneous models, respectively. The plots for |**B**| and $J_n$ show the results at peak transport current, $t = 1.5$ h. The third row show the losses integrated over the cycle.

The losses integrated over the 3 h of the cycle, the losses relative error ($er_q$), the $R^2$, and the computation time are summarized in table 2. The estimated losses by both models, *T-A* multi-scale and *T-A* homogeneous, agree with the *T-A* full model with a relative error equal to 0.28 % and 2.23 %, respectively. Also, both $R^2$ values are higher than 0.95 which demonstrate a good agreement between the *J* distributions. Even though both multi-scale and homogenous models have an acceptable performance, the data in table 2 show that the losses and *J* distributions of the multi-scale model are more accurate.

The most important result is the achieved computation times, 1h 13 min and 37 min, for the multi-scale and homogeneous models, respectively. Both computation times are less than the physical time to cycle the magnet (3 h). Thus, both models are suitable for performing real-time simulations. It is important to mention that the homogeneous model is the less accurate, but at the same time is the fastest. If necessary, the DOF and consequently the computation time can be further reduced for both multi-scale and homogeneous models. The homogeneous model can be reduced by reducing the number of elements along the bulk. The multi-scale model can be reduced, by reducing the number of analyzed tapes, especially in the central pancakes where the losses are lower and do not have a significant impact in the total losses. While it is possible to reduce the DOF in the models, the excessive reduction would compromise the accuracy.

**Table 2.** *T-A* full, multi-scale and homogenous models comparison.

| Model | Losses (J/m) | Relative error (%) | $R^2$ of $J$ | Computation time |
|---|---|---|---|---|
| *T-A* full | 0.8832 | - | - | 9h 04min |
| *T-A* multi-scale | 0.8807 | 0.28 | 0.9888 | 1h 13min |
| *T-A* homogeneous | 0.9025 | 2.18 | 0.9573 | 0h 37min |

To conclude the discussion on the advantages and disadvantages of the different models, it is relevant to mention the difficulty related to the building of the model from scratch in COMSOL. It is not possible to be categorical because it strongly depends on the expertise of the analyst, but some points can be outlined. The models have in common that they require a patient labor to build the geometry and the structured meshes, but this step is particularly easy for the homogeneous model because the number of figures required to build the geometry is considerably lesser than the full models. The *H* full model requires the definition of one integral constraint per tape to impress the transport current, this is a time-consuming process. Conversely, the definition of the boundary conditions that fix the transport current in each tape of the *T-A* full model is a simple procedure that can be done quickly by means of arrays. Once again, this procedure is easier in the homogenous model, because the boundary conditions are defined in the bulks not in the tapes. The *T-A* multi-scale model requires the definition of boundary conditions for the analyzed tapes, but it is necessary to face the additional complication of defining the functions required to implement the *J* interpolation.

## 6. Conclusions

Both the multi-scale and homogeneous techniques have been successfully used with the *H* formulation by various authors. In the present work, we developed these techniques to be used in conjunction with the *T-A* formulation to provide quick and accurate results in large-scale HTS systems.

The *T-A* multi-scale method described in the present work can be seen as enhanced version of the iterative multi-scale method detailed in [35]. The *T-A* formulation allows for the simultaneous computation of **T** and **A** in one single model. Consequently, the *T-A* multi-scale models do not rely on the iterative implementation of several dynamic simulations of two coupled submodels that was the case for the *H* multi-scale model. Additionally, as the *T-A* multi-scale model is mounted on top a single model instead of two submodels, they are easier to build.

The conducted analyzes show that the assumptions considered in the *T-A* homogeneous and multi-scale models significantly reduce the computation time. They led to simplifications in the description of the systems that do not compromise the accuracy of the results. In particular, it allows estimating the current density distribution and subsequently the losses to a high degree of accuracy. The availability of *T-A* homogeneous and multi-scale models open up the possibility for real-time simulation and analysis of large-scale 2G HTS systems. To the best of our knowledge, this capability is new in the analysis of such complex systems, and will bring the opportunity, not only to perform faster parametric simulations, but also to build digital twins of large-scale 2G HTS systems.

The new *T-A* multi-scale and homogeneous approaches are not suitable for the analysis of all the large-scale HTS systems. Since the new approaches rely on the characteristics of the *T-A* formulation, they are suitable for cases where the 1D approximation is meaningful and the influence of the parallel component of the field is negligible. At the same time, these new approaches also inherit the limitations of the multi-scale and the homogeneous models previously described in [20, 33], thus they require less computational resources, but at the price of a lower accuracy than that of the full models. The full models should be applied in the cases where the best possible accuracy is required, if the size of the system and the available computational resources allows it.

## Acknowledgment

We would like to show our gratitude to Magnus Olsson from COMSOL Multiphysics for providing support during the course of this research on the subject of the spurious oscillations.

## Appendix A

During the validation process, it was found that spurious oscillations appeared in the current density computing in the superconducting layers of the HTS tapes. It is hereinafter demonstrated that, with a proper choice of order elements, this issue can be solved. The correct choice of the order of the elements to avoid the spurious oscillations was found by trial and error. Similar issues have already been reported in fluid mechanics literature [43-44], where different order elements are used to approximate velocity and pressure. This issue has not been reported in literature about the *T-A* formulation yet. Until now, the origin of the oscillations is unknown. Interestingly, due to the smoothing and resolution refinement options offered by the postprocessing tools of COMSOL, and sometimes activated by default, it is possible to address these oscillations. In the present work, such tools were not used and bare data were provided.

As discussed in the previous section, the *T-A* formulation uses two state variables, **T** and **A**. Each variable can be approximated using elements of different orders. To assess the correct choice of orders, the *T-A* full model was simulated for three cases of element order combination for one cycle of a 11 A, 50 Hz transport current. The mesh used in the unit cells of the *T-A* full model is structured and consider 60 elements along the tape's width. The first simulation uses first order elements for both **T** and **A**, the second simulation uses second order elements for both variables, while the last simulation uses first order elements for **T** and second order elements for **A**.

The $J$ distributions at the peak transport current ($t = 15$ ms) computed with both $H$ and $T\text{-}A$ full models are shown in figure A1. The normalized current density ($J_n = J/J_c$) for all the tapes is presented in the first row, while the $J$ in tape 96 of pancake 3 is presented in the second row. The results show that the $J$ distributions of the $T\text{-}A$ models present spurious oscillations when the same order of elements is used to approximate both **T** and **A** variables. It is possible to observe that the oscillations are present at subcritical current density values. Also, the period of the oscillations when first order elements are used for both variables is twice the period when second order elements are used for both variables. The oscillations disappear when first order elements are used for **T** and second order elements for **A**.

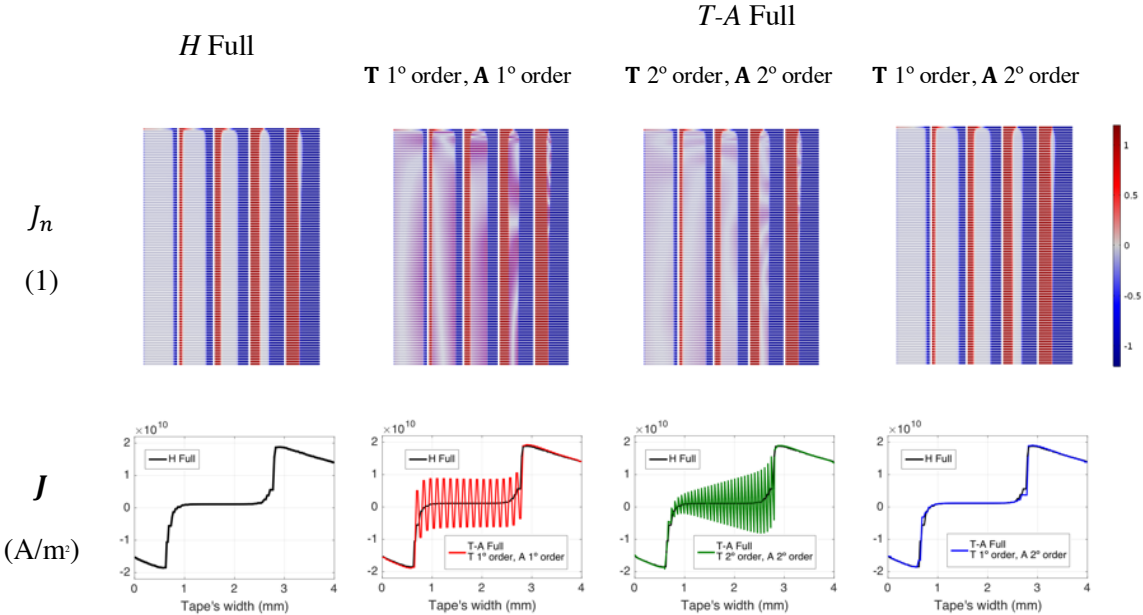

**Figure A1.** $J$ distribution at peak transport current ($t = 15$ ms) for $H$ full and $T\text{-}A$ full models. The first rows show $J_n$ for all the tapes. The second row shows $J$ in the tape 96 of pancake 3. The simulations were performed with the $H$ formulation reference model and with the $T\text{-}A$ full model. The $T\text{-}A$ full model uses different order elements for **T** and **A**. The spurious oscillations are present when the same order is used for both variables. No refinement and smoothing were used in the postprocessing.

Table A1. *H* and *T-A* full models comparison.

| Model | | Av. Losses (W/m) | Relative error (%) | $R^2$ of $J$ | Computation time |
|---|---|---|---|---|---|
| | *H* full | 127.2389 | | | 31h 32min |
| *T-A* full | **T** 1°, **A** 1° order | 126.0212 | 0.96 | 0.8905 | 02h 46min |
| | **T** 2°, **A** 2° order | 126.3897 | 0.67 | 0.9884 | 09h 01min |
| | **T** 1°, **A** 2° order | 128.0560 | 0.64 | 0.9922 | 03h 14 min |

The average losses, the losses relative error ($er_q$), the $R^2$, and the computation time are summarized in table A1. The spurious oscillations are present at subcritical values, so they have a negligible impact on the hysteresis losses. The estimated losses of the *T-A* models agree with the *H* full model with less than 1 % error, but the lower relative error is achieved with the model using first order elements for **T** and second order elements for **A**. The impact of the oscillations is clearly reflected in the $R^2$ values that correspond to the goodness of the fit of the *J* distribution. The smaller $R^2$ value of the model using second order elements for both variables compared to the model using first order elements reflects larger amplitudes in the oscillations as shown in figure A1. The model with the largest $R^2$ value is the model using first order elements for **T** and second order elements for **A**, the one without oscillations.

The three *T-A* full models have a computation time significantly lower than the *H* full model. The increment in the order of the elements leads to the increment in the degrees of freedom (DOF), and the computation time, then the computation time when second order elements are used for both variables is almost three times the computation time of the model using first and second order elements. It should be mentioned that with the current version 5.3 of COMSOL, it was not possible to run simulations using second order elements for **T** and first order elements for **A**.

## Appendix B

The models using the *H* formulation [27] are more complex for simulating large-scale HTS systems under slow dynamic conditions. Indeed, it was found that the computation time increases as the frequency decreases. Different solvers and solver settings of COMSOL have been tried out without success in order to decrease the computation time. It should be acknowledged that COMSOL is a proprietary software for which we do not have a complete access to the solver settings. Conversely, the models using the *T-A* formulation have been fairly simple to use and successful for simulating slow cycles, while keeping reasonable computation times.

Here the *benchmark #3* of the *HTS modelling workgroup* [45] was used to study the increase in the computation time when the frequency is gradually reduced. The *benchmark 3#* is a 20-tape stack representative of a racetrack coil. The parameters of the *benchmark 3#* are resumed in table B1. Two models were built, the first model is a *H* formulation model and the second one is a *T-A* formulation model. Both models consider each individual tape, as in the *H* full and *T-A* full models. The mesh used in both models is structured and consider 60 elements along the tape's width.

**Table B1.** Benchmark 3# parameters.

| Parameter | Value |
|---|---|
| Number of tapes | 20 |
| Inner radius | 10 cm |
| HTS layer width | 12 mm |
| HTS layer thickness | 1 μm |
| Distance between tapes | 250 μm |
| $I_c$ | 300 A |
| $N$ | 25 |

The considered transport currents are sinusoidal with amplitude $0.5*I_c = 150$ A, and frequencies ranging from 5e-5 Hz to 50 Hz. The hysteresis losses per cycle and the computation time as a function of frequency are shown figure B1 and B2, respectively. As a side note, the results in figure B1 show that the losses per cycle have a small frequency dependence, similar results have been reported in [46-50].

Figure B2 shows that the computation time required to simulate one sinusoidal cycle using the *T-A* full model is independent of the frequency, and remains around 300 s for all the frequency range. On the contrary, the computation time required by the *H* full model expands up to 36.6 h when the frequency is 5e-5 Hz. Only the lower computation time has been presented for each case. It should be emphasized that the computation times of the *H* full model reported in figure B2 are in accordance with the practical experience of the authors.

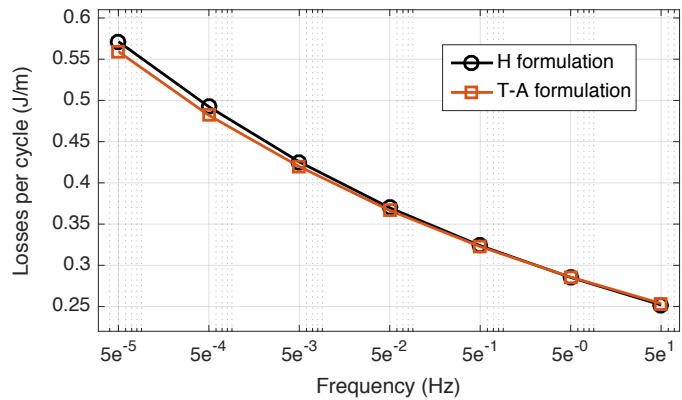

**Figure B1.** Benchmark #3. Losses per cycle as a function of frequency.

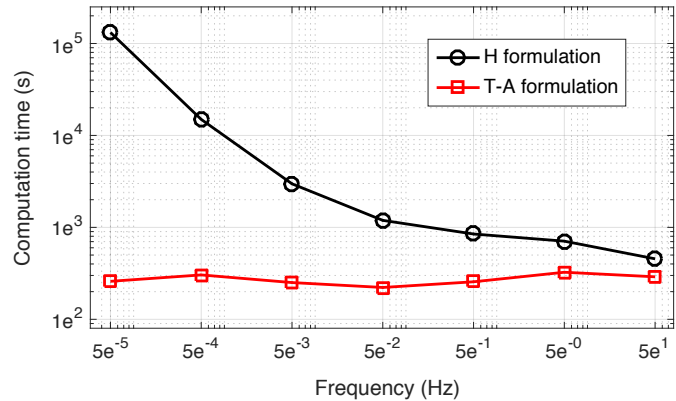

**Figure B2.** Benchmark #3. Computation time required to simulate one period of the sinusoidal transport current as a function of frequency.